\documentclass[journal]{IEEEtran}

\usepackage{amssymb,amsmath,epsfig,booktabs,tipa,array,multirow,url}

\begin{document}
%
\title{Objective Study of Sensor Relevance for Automatic Cough Detection}

\author{Thomas Drugman, Jerome Urbain, Nathalie Bauwens, Ricardo Chessini, Carlos Valderrama, Patrick Lebecque and Thierry Dutoit
\thanks{Thomas Drugman, Jerome Urbain, Ricardo Chessini, Carlos Valderrama and Thierry Dutoit are with the University of Mons, Belgium.}
\thanks{Nathalie Bauwens and Patrick Lebecque are with the Pediatric Pulmonology \& Cystic Fibrosis Unit, Cliniques St Luc, University of Louvain, Belgium.}}

\maketitle

\begin{abstract}
The development of a system for the automatic, objective and reliable detection of cough events is a need underlined by the medical literature for years. The benefit of such a tool is clear as it would allow the assessment of pathology severity in chronic cough diseases. Even though some approaches have recently reported solutions achieving this task with a relative success, there is still no standardization about the method to adopt or the sensors to use. The goal of this paper is to study objectively the performance of several sensors for cough detection: ECG, thermistor, chest belt, accelerometer, contact and audio microphones. Experiments are carried out on a database of 32 healthy subjects producing, in a confined room and in three situations, voluntary cough at various volumes as well as other event categories which can possibly lead to some detection errors: background noise, forced expiration, throat clearing, speech and laugh. The relevance of each sensor is evaluated at three stages: mutual information conveyed by the features, ability to discriminate at the frame level cough from these latter other sources of ambiguity, and ability to detect cough events. In this latter experiment, with both an averaged sensitivity and specificity of about 94.5\%, the proposed approach is shown to clearly outperform the commercial Karmelsonix system which achieved a specificity of 95.3\% and a sensitivity of 64.9\%.


\end{abstract}


\IEEEpeerreviewmaketitle

\section{Introduction}\label{sec:intro}

\IEEEPARstart{C}{ough} is defined in physiology textbooks as a three-phase act but also, for clinical purposes, as an expiratory manoeuvre against a closed glottis, which produces a characteristic sound. It is a protective mechanism of the proximal respiratory tract. It may be voluntary or reflex.  Cough can appear sporadically with common illnesses (e.g. cold), but when it becomes chronic it can severely impair life quality. This symptom is the commonest reason for which people seek medical advice\cite{Schappert}. It concerns one third of pulmonologist consultations\cite{McGarvey}. The assessment of cough severity is an essential tool in clinical use. It requires a combination of measures characterizing cough frequency, intensity and its impact on quality of life.
 
The evaluation of cough severity is currently subjective: it is based on cough scores, diaries, visual analogue scales (VAS) and symptom questionnaires, which are completed either by the patient himself or a parent \cite{Birring} \cite{Marsden}. However, it has been shown that subjective assessments correlate modestly with objective measures of cough frequency \cite{Decalmer}. Medical literature underlines the lack of an objective and reliable tool to measure the severity of this symptom \cite{ERJ}. Such a tool is also necessary to test the effectiveness of cough treatments (e.g. antitussives) or novel therapies.

Cough events can be counted manually on audio recordings but this task remains extremely laborious and time-consuming, and hence economically unfeasible. Recent technological advances have enabled the development of automated and ambulatory cough monitors. Existing systems use audio signal(s) alone (microphone and/or contact microphone) or in combination with other sensors such as accelerometers, pneumographic belt, electromyography electrodes, electrocardiography electrodes, induction plethysmography and pulse oxymeter. However, as it is underlined by the European Respiratory Society committee, there is currently neither standardized methods for recording cough nor adequately validated, commercially available and clinically acceptable cough monitors \cite{ERJ}.
 
Automatic cough monitors have to face several challenges. First, cough can be confused with surrounding noise and other parasitic patient sounds (throat clearing, laughing, forced expiration, etc.). Secondly, the acoustic properties of cough events vary with individuals and diseases, but also within individuals. Finally, it is desirable to monitor cough continuously for long periods (over 24h) and in the patient's own environment. 

A few cough monitors have been developed so far \cite{Smith1}. It started around the 1950s with simple audio recording systems enabling to manually spot the cough events. It is only recently that (semi- or fully-) automated cough recorders have been designed.  

Some cough monitors use the audio signal only. In 2006, the Hull Automated Cough Counter ($HACC$) \cite{Barry} was developed. It consists of a single audio signal fed into an artificial neural network for detecting cough events. The number of cough components per event is not computed by the system, but can be manually determined. The system presents a sensitivity of 80\% (ranging from 55\% to 100\% accross the 10 validation patients) and a specificity of 96\%. The Leicester Cough Monitor ($LCM$) \cite{LCM} also relies only on audio recordings. It enables 24h ambulatory recordings. Hidden Markov Models are used to pre-segment possible cough events. Some of these possible cough segments are then presented to a human expert in order to develop a statistical model tailored to the current recording. Finally, the full recordings are processed with the developed models, achieving overall sensitivity and specificity of 91 and 99\%, respectively. Matos et al. \cite{Matos06} also use Hidden Markov Models trained on audio features only, with an approach inspired by keyword spotting: models are developed to characterize cough events, but also to represent the set of all other possible events (named ``fillers''). The two models compete to score new recordings and the most likely sequence of coughs and fillers is retained. This system achieves a sensitivity of 78\%, ranging from 50\% to 99\% depending on the subjects, with false alarm rates ranging from 2 to 31 per hour. The performance is lower than the $LCM$ but here the process is fully automatic. The Vitalojak system \cite{Smith2} uses only a contact microphone placed on the thorax for a semi-automated detection. A lapel microphone is added to the system for manual validation. This system has been validated in a 24-hour ambulatory context on 10 patients \cite{McGuinness}. It reaches a sensitivity higher than 99\% while compressing the amount of data to check manually with 3 possible levels, ranging from 65 to 23 minutes on average. In \cite{Drugman-IS12}, we proposed an audio-only system based on artificial neural networks which achieved a sensitivity and specificity of about 95\% on voluntary cough from 10 healthy subjects in various conditions.



Two commercial systems rely on multiple signals to detect cough. In 2005 appeared the Lifeshirt \cite{Coyle}, consisting of several sensors integrated in a shirt worn by the user: electrocardiogram, induction plethysmography, 3-axis accelerometer and a contact microphone placed on the throat. The device achieved sensitivity and specificity of 78.1 and 99.6\%, respectively. It must be noted that the Lifeshirt is no longer available due to the liquidation of the company in 2009. In 2010, the Karmelsonix company launched the PulmoTrack-CC\textsuperscript{TM} \cite{Vizel}, which includes a piezoelectric belt for monitoring respiration, one lapel microphone and two contact microphones placed on the trachea and the thorax. The performance of this device is reported in \cite{Vizel} to reach a sensitivity of 91\% and a specificity of 94\% on voluntary cough. Although it has not been clinically validated, it is currently commercialized and will be used as a comparison point in our experiments.

In this paper we will present the first development steps towards a new, reliable, ambulatory cough monitor, following the recommendations of the European Respiratory Society. The paper focuses on the selection of sensors and the development of algorithms to provide increased recording capabilities, detection rates and robustness (i.e. efficiency for all the subjects) compared to the existing systems. Clinical validation will be performed in the near future. The paper is organized as follows: the acquisition system is presented in Section \ref{sec:Acqui}; the database recorded to develop and evaluate the cough monitor is presented in Section \ref{sec:database}; the developed framework for cough detection is detailed in Section \ref{sec:Framework}; results of feature selection are described in Section \ref{sec:ResultSelection}; Section \ref{ssec:ResultCategory} focuses on the discrimination of cough with regard to other event categories (background, forced expiration, throat clearing, speech and laugh); in Section \ref{sec:Karmelsonix} finally a comparison between the proposed approach and the commercially available Karmelsonix system is carried out; finally Section \ref{sec:Conclusion} draws the conclusions of this study.

\section{Acquisition System}\label{sec:Acqui}

The acquisition system aims at capturing and storing data from sensors. Signals are not analyzed or processed along this step, which is performed off-line (as explained in Section \ref{sec:Framework}). The block diagram of the acquisition system is displayed in Figure \ref{fig:AcquisitionSystem}. It is implemented as a classical acquisition system by means of sensors, analog signal conditioning circuit (front-end), analog-to-digital conversion, communication, and storing functional blocks. The general specifications of the acquisition system are the following:

\begin{figure}[htbp]
  \begin{center}
   \includegraphics[width=7.5cm]{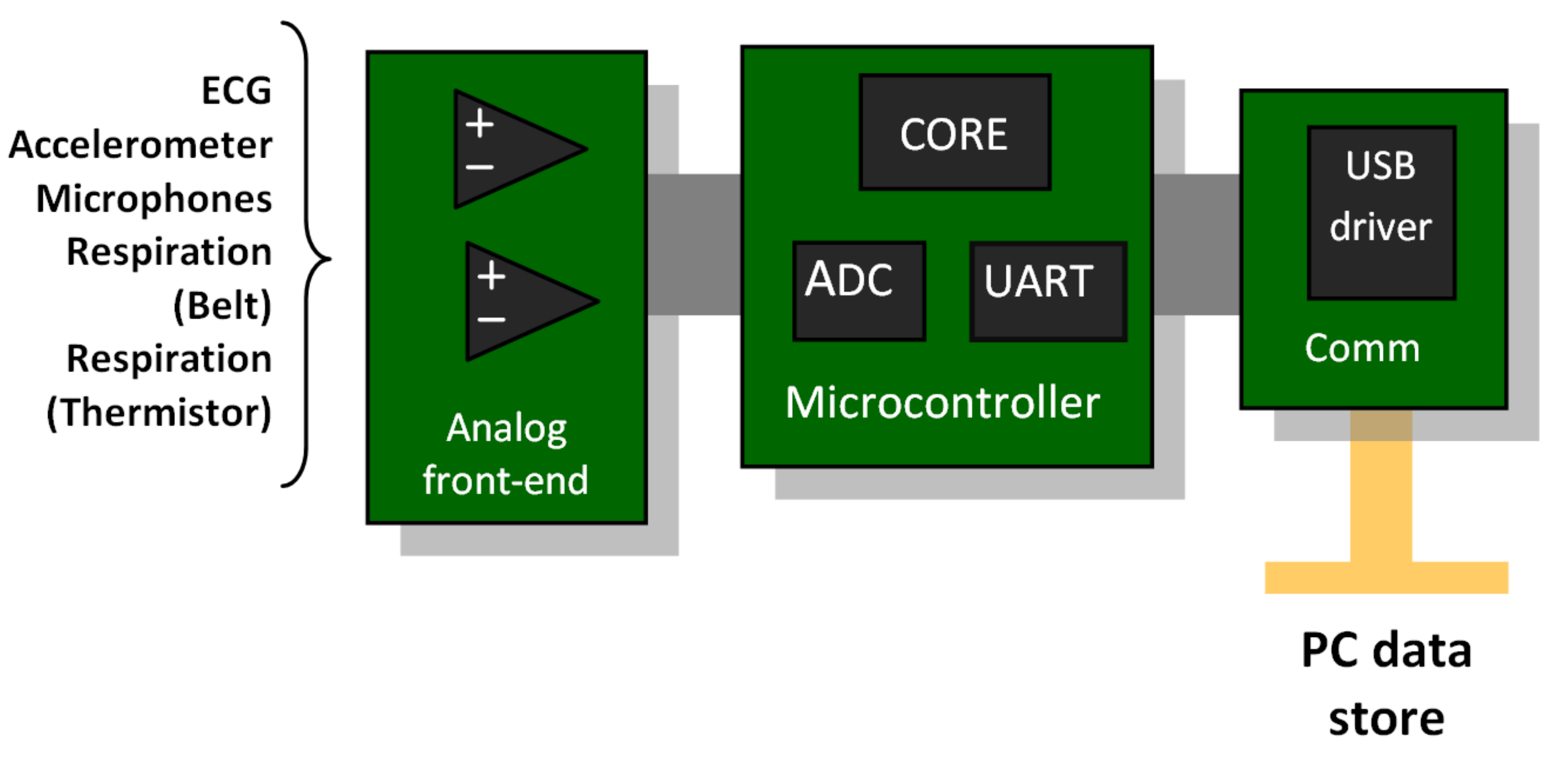}
  \end{center}
  \vspace{-.5cm}
\caption{{\it Block diagram of the acquisition system.}} 
\vspace{-.2cm}
\label{fig:AcquisitionSystem}
\end{figure}

\begin{itemize}
 \item
\textbf{Sensors}: Six sensors are used in this study: \emph{i)} a disposable ECG monitoring electrodes 40x36x1 mm, \emph{ii)} a SLP nasal thermocouple flow sensor from Medatec (specification not available) measuring the thermical flow at the patient's nose, \emph{iii)} a piezo respiratory effort belt, \emph{iv)} an ADXL103 analog dual axis accelerometer from \textit{Analog Device} with $\pm$1.7g/$\pm$5g/$\pm$8g of precision and $\pm$1mg of resolution, \emph{v)} a contact omni-directional microphone placed on the throat, with 42dB of signal-to-noise ratio and 40-12000Hz of bandwidth, and with 2mm of silicone cup, and \emph{vi)} a lapel audio electrets omni-directional microphone featuring 60dB of signal-to-noise ratio and 50-16000Hz of bandwidth.

\item
\textbf{Analog front-end}: Each sensor has an amplifier circuit based on an AD620 instrumentation amplifier from \textit{Analog Device} featuring 100dB of common-mode rejection, 50$\mu$V max input off-set and 120KHz of bandwidth. The analog circuits are supplied by a dedicated symmetric power of 5 volts. Each sensor circuit was designed to adjust the signal between 0 to 2.6 V of amplitude and no more than 10 KHz of frequency as requirements. 

 \item
\textbf{Analog-to-digital conversion}: This step is performed by 8-channel and 12-bit ADC as part of ADuC841 microcontroller from \textit{Analog Device}. For the two microphones, acquisition is made at 10000 samples per second. For other sensors, since their signal (which we call \textit{biosignals} in opposition to the acoustic signals captured by the microphones) have a much slower dynamics, the sampling rate is fixed to 1250 Hz. 

 \item
\textbf{Communication}: This step is based on a USB communication achieved by FT232R USB-to-serial component from FTDI, with intermediate serial UART communication from ADuC841 and USB communication from a PC on 115200 Bits/s of data rate.  

 \item
\textbf{Storing}: A software in C++ manipulates the data from USB and stores it as {\emph{.raw}} files in the PC hard disk. 
\end{itemize}

The key idea here is to be as exhaustive as possible by analyzing the performance of several sensors. Some of them, like the audio and contact microphones, the accelerometer or the chest belt, have been used in various existing cough monitors \cite{Smith2}. Some others (e.g. the nasal thermistor, or the ECG) are less common for cough detection, and their inclusion in this study is more exploratory by investigating their potential comparatively to more usual sensors.

\begin{figure}[htbp]
  \begin{center}
   \includegraphics[width=8cm]{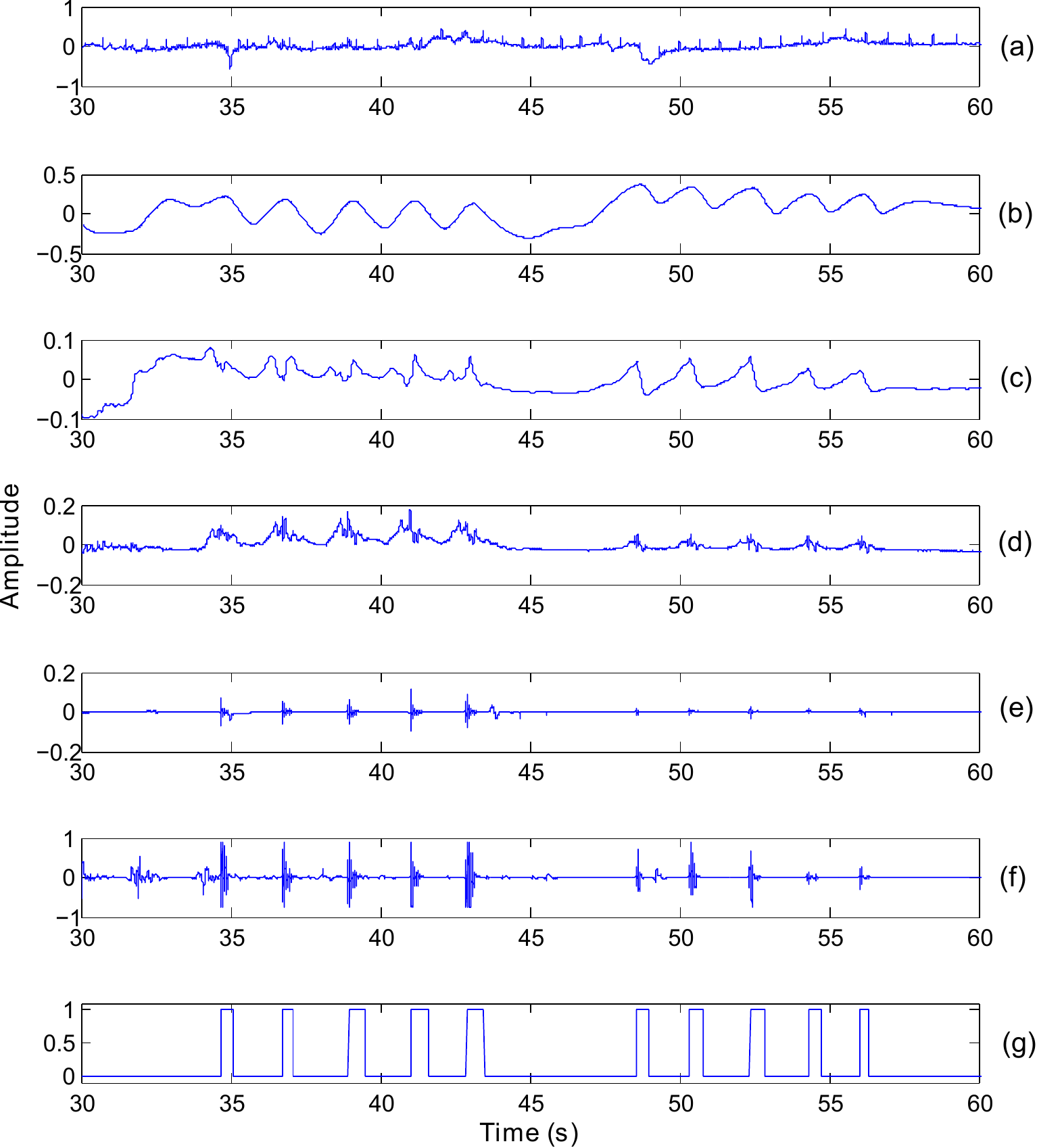}
  \end{center}
  \vspace{-.5cm}
\caption{{\it Example of signals captured by the various sensors: a) ECG, b) thermistor, c) chest belt, d) accelerometer, e) contact microphone, f) audio microphone, and g) manual cough event label.}} 
\vspace{-.2cm}
\label{fig:Plot_Signals}
\end{figure}

Figure \ref{fig:Plot_Signals} illustrates the signals recorded from these sensors by the acquisition system during a period of coughing (as indicated by the manual annotations in the last panel). Heartbeat pulses can be seen on the ECG recording (panel (a)). This signal can exhibit some artifacts during cough events which can be turned into advantage for their detection (such as a bit before 35s in panel (a)). As observed in panels (b) and (c), the signals captured from the thermistor and the chest belt are mainly informative about the respiratory volume. They are both very slow-varying and show local maximum values around the cough instants, characterized by important expiratory flow and respiratory volume variations. The signal from the accelerometer (panel (d)) is clearly observed to present a much higher activity in the coughing neighborhood due to the abdominal contractions in such periods. Finally the signals acquired by the contact and audio microphones are shown in panels (e) and (f) respectively. Although the contact microphone is more robust to ambient noise, it suffers from other acquisition perturbations, notably due to the fact that with such a sensor, it is difficult in some conditions (subject moving, etc) to maintain a good contact with the skin, leading to some degradation during the acquisition (albeit it is not the case in Figure \ref{fig:Plot_Signals}). The use of a contact microphone with better attachment might alleviate these robustness issues.

\section{Database}\label{sec:database}
The study population was divided into two groups. The first set (A) included 22 healthy subjects (9 Male, mean age$\pm$SD: $22.8\pm2.44$ ,range: $20-28$). The second set (B), consisting of 10 additional healthy subjects (5 Male, mean age$\pm$SD: $23\pm1.45$, range: $22-26$) was designed to compare our system to the commercially available KarmelSonix cough counter \cite{Vizel}. It is worth noting that these recordings were made across several sessions and in different rooms.


\begin{table} 
  \begin{center}
\vspace{0cm}
    \begin{tabular}{|p{1.8cm}|p{1.8cm}|p{1.8cm}|p{1.8cm}|}
\hline
\multirow{3}{*}{\textbf{Sounds}} & \multicolumn{3}{c}{\textbf{Situations}}\\\cline{2-4}
& Sitting down, quiet environment (n) & Sitting down, noisy environment (n) & Climbing/going down a stepladder (n) \\
\hline
High volume cough & 5 & 5 & 5\\
Interm. vol. cough & 5 & 5 & 5\\
Low vol. cough & 5 & 5& 5\\
Fit of coughing & 3 & 3 & 3\\
Forced expiration & 3 & 3 & 3\\
Throat clearing & 5 & 5 & 5\\
Speaking & 14 & 14 & 14\\
Laughing & 3 & 3 & 3\\
\hline
    \end{tabular}
  \end{center}
\vspace{-.2cm}  
  \caption{\emph{Standardized protocol for data recorings}}
\vspace{-.8cm}
\label{tab:database_protocol}
\end{table}


The aim of the database was to record various cough sounds but also some other sounds which are typically confused with cough. The participants followed a standardized protocol performed in three different situations, as detailed in Table \ref{tab:database_protocol}: (a) sitting down in a quiet environment, (b) sitting down in a noisy environment and (c) climbing on/going down of a stepladder. 

This protocol was inspired by the one used to develop and evaluate the PulmoTrack-CC\textsuperscript{TM} \cite{Vizel}, with the addition of low and intermediate volume coughs as these kinds of cough are more difficult to detect for automatic cough counters.
All recordings have been precisely manually annotated by a trained observer. In total, the database contains 2338 coughs (among which 864 are from fits of coughing), 289 forced expirations, 479 throat clearings, 289 laughters, for a total duration of 237 minutes. Note that slight deviations were observed from the strict protocol, but that the manual annotation was made coherently.

\section{Cough Detection Framework}\label{sec:Framework}

\begin{figure}
  \begin{center}
   \includegraphics[width=8cm]{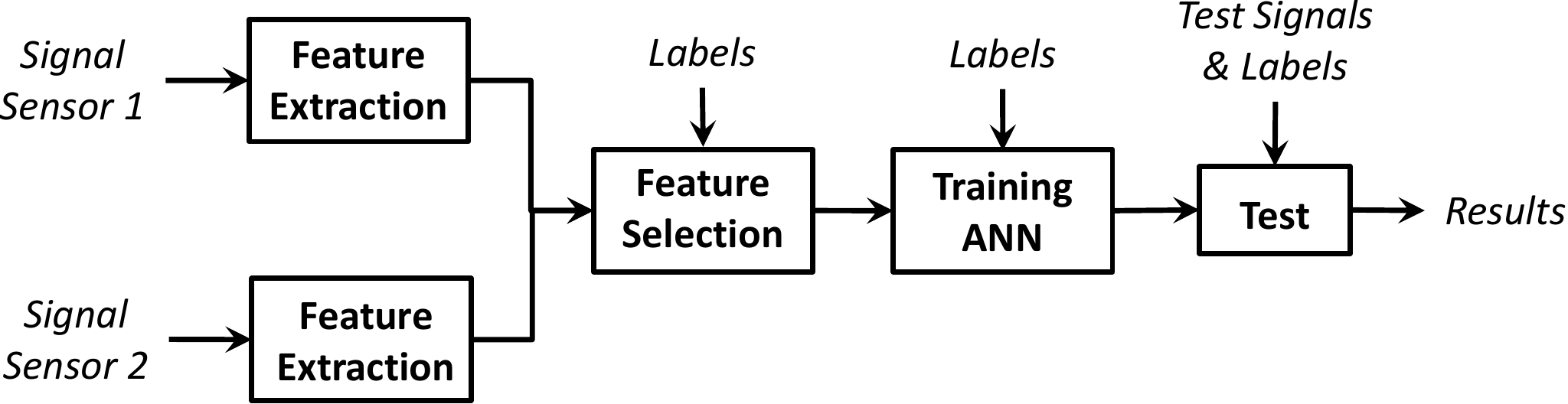}
  \end{center}
  \vspace{-.5cm}
\caption{{\it Workflow used in this study for the automatic multimodal detection of cough.}} 
\vspace{-.2cm}
\label{fig:Workflow}
\end{figure}

The general workflow for the automatic detection of cough used troughout this paper is displayed in Figure \ref{fig:Workflow}. From the signals captured by a given sensor, or by several sensors in a multimodal approach, the first step aims at extracting a wide variety of features, as described in Section \ref{ssec:Extraction}. Since this leads to a prohibitive number of features, a step of dimensionality reduction is necessary by selecting only the most relevant ones (Section \ref{ssec:Selection}). This is here achieved based on some measures derived from the Information Theory. Finally, an Artificial Neural Network (ANN) based classifier is used for accurately modeling the feature distributions and for drawing the final cough detection decision (Section \ref{ssec:ANN}).

\subsection{Feature Extraction}\label{ssec:Extraction}

Many features, both in the time and frequency domains, have been proposed in the literature to describe the characteristics of a signal. The key idea here is to extract the largest variety of features among which only the most relevant will be selected (Section \ref{ssec:Selection}). In other words, we target at a comprehensive description of the signals from which data mining techniques will retain only the most relevant for our classification problem.

Features used in this study can be categorized into 3 classes:

\begin{itemize}

\item{\textbf{Features Describing the Waveform}}
\newline
These descriptors are simply extracted from the signal waveform in the time domain. They consist of the signal amplitude at the analysis time as well as the signal slope computed at different scales ranging from 1.28s to 0.04s. These features might be of interest for slow-varying signals, for example for the thermistor or the chest belt which capture respiratory characteristics. Unlike the next two categories, they are in the following extracted only for biosignals, and are therefore not considered for the two microphones for which they are not appropriate.

\item{\textbf{Features Describing the Spectral Contents}}
\newline
Several features characterizing the spectral contents have been proposed in \cite{Peeters}. For a comprehensive description of the magnitude spectrum, we extracted the three following types of descriptors: the Mel-Frequency Cepstral Coefficients (\textit{MFCCs}, \cite{MFCC}) as widely used in automatic speech recognition, the specific \textit{loudness} associated to each Bark band \cite{Loudness} and the \textit{relative energy} in different frequency subbands linearly spaced by 500 Hz. In order to be as exhaustive as possible, we also used the \textit{chroma} features \cite{Chroma} as widely done in music processing, although their potential might be limited for our issue.

Besides, several parameters describing the spectral shape are also employed, as proposed in \cite{Peeters}. The \emph{Spectral Centroid} is defined as the barycenter of the amplitude spectrum. Similarly, the \emph{Spectral Spread} is the dispersion of the spectrum around its mean value. The \emph{Spectral Decrease} is a perceptual measure quantifying the amount of decreasing of the spectral amplitude \cite{Peeters}. The \emph{Spectral Variation} and \emph{Spectral Flux} characterize the amount of variations of spectrum along time and are based on the normalized cross-correlation between two successive amplitude spectra \cite{Peeters}. Finally, we also use the \textit{energy} and \textit{total loudness} which are mainly informative about the presence of activity in the signal.

\item{\textbf{Features Measuring the Amount of Noise}}
\newline
Quantifying the level of noise may be of interest for describing the cough sound for audio signals, or for detecting movements from accelerometer recordings. For this, several measures are here extracted. First, the \emph{Harmonic to Noise Ratio} (HNR) is calculated in the four frequency ranges [0-0.5kHz], [0-1.5kHz], [0-2.5kHz] and [0-3.5kHz]. The \emph{Spectral Flatness} measures the noisiness/sinusoidality of a spectrum (or a part of it) in four frequency bands \cite{Peeters}. The \emph{Zero-Crossing Rate} quantifies the number of times the signal crosses the zero axis. It is expected that the greater the amount of noise, the higher the number of zero-crossings. The $F_0$ value and its related measure of periodicity based on the Summation of Residual Harmonics \cite{SRH} are used as voicing measurements. As a last parameter quantifying the amount of noise in the audio signal, the \emph{Chirp Group Delay} is a phase-based measure proposed in \cite{DrugmanPhase} and shown to be suited for highlighting turbulences in the signal.
\end{itemize}

These features are extracted every 12 ms for all sensors so as to have synchronous streams. As mentioned above, the first feature category was only extracted for biosignals, while the two other categories were used in all cases. For microphones, the window length was fixed to 30 ms, a standard value in audio processing, while two frame sizes were used for the biosignals: 500 ms and 1 s. Finally, we also added the first and second derivatives for each of these features in order to integrate the signal dynamics.

\subsection{Feature Selection}\label{ssec:Selection}
A very large number of features has been extracted in Section \ref{ssec:Extraction}: 222 for the two microphones, and 369 for the biosignals (including the first and second derivatives). The goal of the feature selection algorithm is to retain the most relevant ones so as to alleviate the effect of the curse of dimensionality \cite{Bellman}. For this, we here make use of measures based on the mutual information. This allows to assess the intrinsic discrimination power of each feature separately, but also their possible complementarity or redundancy, and this independently of the subsequent classifier.

The feature selection algorithm used in this study has been proposed in \cite{Drugman-FS}. It is a greedy method which at each iteration chooses the feature conveying the greatest amount of new relevant information for the considered classification problem. This latter measure is estimated by considering the mutual information of this feature and its redundancy with the selected subset.

In the rest of the paper, only 20 features are used for each configuration (i.e for a particular signal or for a possible combination of several sensors). The relevance of these features is discussed throughout Section \ref{sec:ResultSelection}.

\subsection{Artificial Neural Network-based Classification}\label{ssec:ANN}

For each of the issues tackled across experiments, a dedicated Artifical Neural Network (ANN) has been trained. Our ANN implementation relies on the Matlab Neural Network toolbox. Each ANN is made of a single hidden layer consisting of neurons (fixed to 32 neurons in this work) whose activation function is an hyperbolic tangent sigmoid transfer function. The output layer is a simple neuron with a logarithmic sigmoid function suited for a binary decision.

In order to provide contextual information to the ANNs, the feature vector at the considered analysis time is appended with its values 50 ms and 100 ms both in the past and in the future. Finally note that when testing, the posterior probability of cough detection, provided by the ANN, is smoothed by a median filtering over a period of 50ms so as to remove erroneous isolated decisions.


\section{Results of Feature Selection}\label{sec:ResultSelection}

The various features for each sensor are first evaluated based on their mutual information with the classes, i.e. on their discrimination power for detecting cough. The computation of mutual information requires to estimate probability density functions of the features jointly with the classes. This is here carried out via a histogram approach \cite{Drugman-FS}. The number of bins is set to 50 for each feature dimension, which results in a trade-off between an adequately high number for an accurate estimation, while keeping sufficient samples per bin. Class labels correspond to the presence (or not) of a cough event. These measures are then normalized over the class entropy. This ensures a more intuitive interpretation with values ranging between 0 (the feature does not convey any relevant information) and 1 (the feature covers 100\% of the classification information).

\begin{figure}[htbp]
  \begin{center}
   \includegraphics[width=8cm]{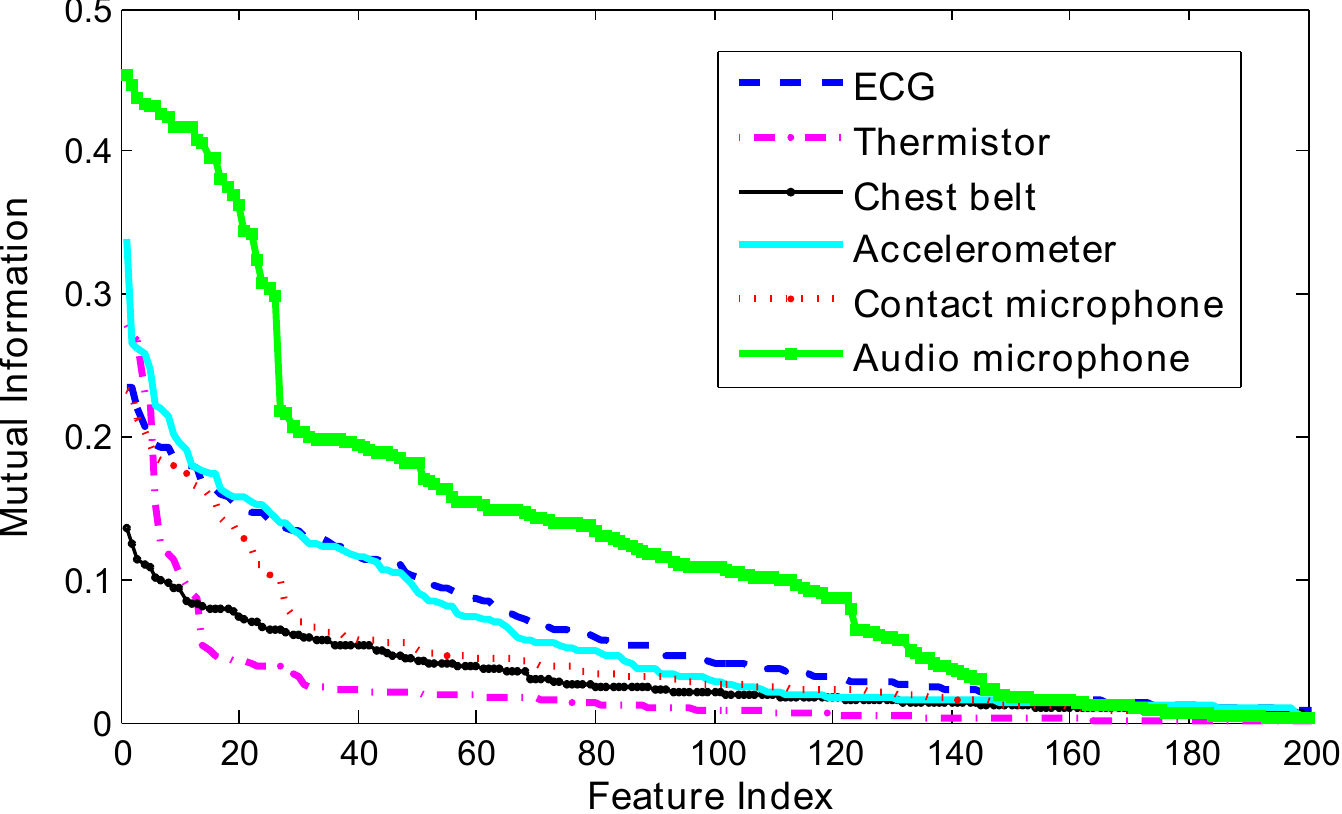}
  \end{center}
  \vspace{-.5cm}
\caption{{\it Mutual Information between a feature (considered independetly) and the classes, for the 200 best features and for all sensors.}} 
\vspace{-.2cm}
\label{fig:MutualInfo}
\end{figure}

Mutual information measures are exhibited in Figure \ref{fig:MutualInfo} for the best 200 features \textit{considered independently} for each sensor (therefore their redundancy is not taken into account). It is clearly seen that the audio microphones provides the largest amount of relevant features. It is followed by the accelerometer and the ECG which seem to give a similar amount of information. It turns out for the thermistor that a limited number of features (about 15) bring an interesting amount of information, while its other features are poorly informative. Finally, the worst sensor is observed to be the chest belt.

\begin{figure}[htbp]
  \begin{center}
   \includegraphics[width=8cm]{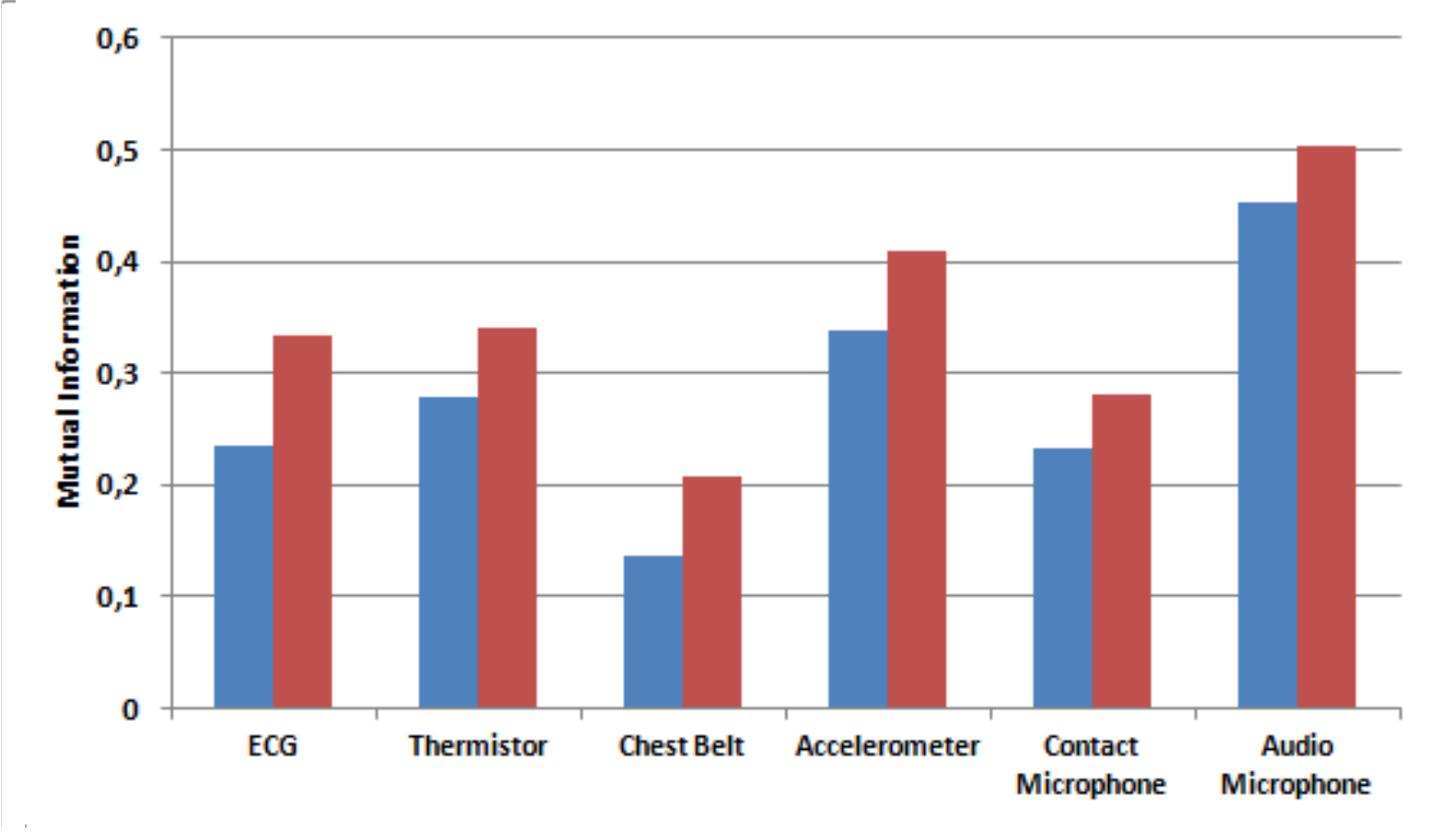}
  \end{center}
  \vspace{-.5cm}
\caption{{\it Mutual information for the best feature (in blue), and 2 best features (considered jointly - in red) of each sensor.}} 
\vspace{-.2cm}
\label{fig:FeatSelect}
\end{figure}

Figure \ref{fig:FeatSelect} shows the mutual information for the best feature and the best 2 features considered jointly, i.e. taking their redundancy/synergy into account. This study of redundancy is only possible for two features since investigating more features would require to consider probability functions in spaces of high dimension, where they cannot be estimated accurately \cite{Bellman}, \cite{Drugman-FS}. With only two features, the best sensors are observed to be: the audio microphone (0.50), the accelerometer (0.41), the thermistor and the ECG (with both about 0.34), the contact microphone (0.26) and finally the chest belt turns out to be the worst sensor (0.20).

The most interesting features for each sensor, which were chosen by the feature selection algorithm, are: \textit{i)} for the ECG, the specific loudness in the fourth and fifth Bark bands (from 204 to 399 Hz)) and their derivatives, \textit{ii)} for the thermistor, the signal slopes at various scales (from 160ms to 1.28s) and the signal amplitude at the analysis time, \textit{iii)} for the chest belt, the energy over a window of 500ms, and the signal slopes at various scales (from 160 to 640 ms), \textit{iv)} for the accelerometer, the specific loudness in the second and fourth Bark bands (from 13 to 297 Hz) and their derivatives, and the energy in a 1s-long window, \textit{v)} for the contact microphone, the specific loudness in the Bark bands 7, 11, 12 and 19 (respectively [631-765],[1268-1479],[1479-1720] and [4386-5000] Hz), \textit{vi)} and for the audio microphone, the total loudness and its derivative, as well as the first derivative of the specific loudness in the Bark bands 18 and 19 (from 3702 to 5000 Hz).


\section{Discriminating Cough from Other Event Categories}\label{ssec:ResultCategory}

\begin{figure*}[htbp]
  \begin{center}
   \includegraphics[width=17cm]{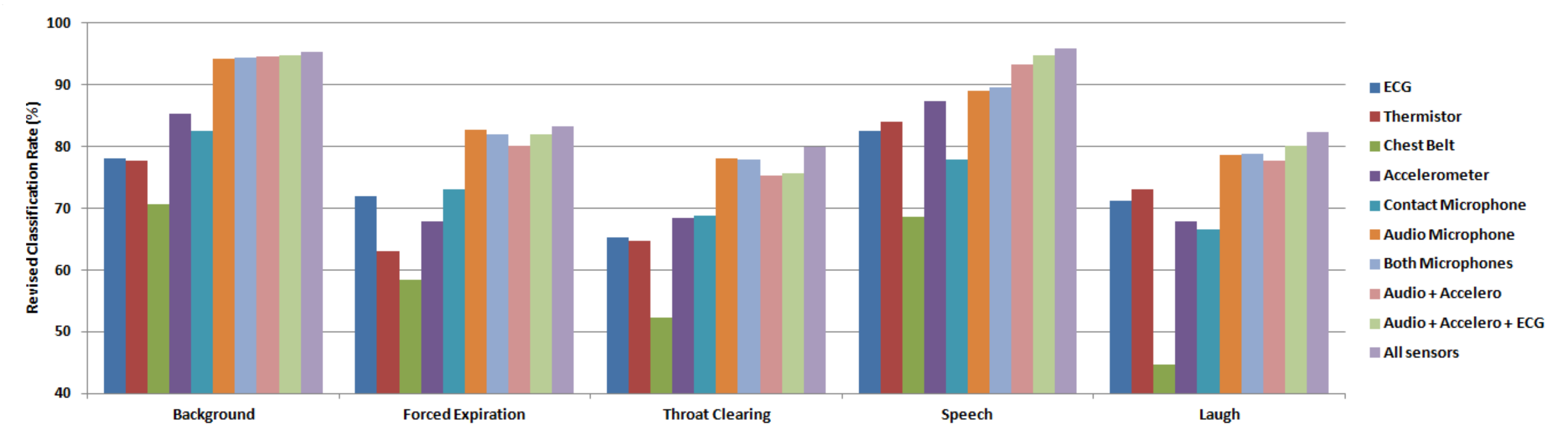}
  \end{center}
  \vspace{-.5cm}
\caption{{\it For each of the problems discriminating cough with another event category (indicated on the horizontal axis), the Revised Classification Rate is given at the frame level using various sensors and their possible combinations.}} 
\vspace{-.2cm}
\label{fig:CategoryResults}
\end{figure*}

Several event categories can be sources of ambiguity for the final cough detection issue, and the performance of a given sensor may vary across these categories. For example, it might turn out that a sensor discriminates well cough from laugh but has a poor performance for the differentiation with throat clearing. We therefore focus in this section on the detection of cough with regard to various other categories of events: background, forced expiration, throat clearing, speech and laugh.

For this experiment, we only consider segments of the whole database (sets A and B) containing either cough or the considered category. For each one of these two categories, a specific ANN is trained. Performance is evaluated at the frame level using a leave-four--subjects-out cross-validation procedure. In other words, 28 out of the 32 subjects are used for the training and the four remaining are used for the test. This operation is repeated 8 times so as to cover the whole database for the test.

Metrics we use at the frame level are the True Positive Rate (TPR) and False Positive Rate (FPR). By varying the threshold decision $\theta$ (applied on the posterior probability outputted by the ANN), a Receiver Operating Characteristic (ROC) curve is obtained in the (FPR,TPR) plane \cite{Hanley}. As a single measure of performance at the frame level, we use the Revised Classification Rate (RCR), defined as:

\begin{equation}\label{eq:RCR}
RCR= 1 - \min_{\theta} \sqrt{(1-TPR(\theta))^2+FPR(\theta)^2}.
\end{equation}

Indeed, an ideal classifier being characterized by a $TPR=1$ and a $FPR=0$, a single measure of performance is the Revised Error Rate ($RER=1-RCR$) defined as the Euclidian distance from the top-left corner to the ROC curve. As a consequence, the higher RCR (the lower RER) the better the system. This criterion implies that an equal importance is given to both TPR and FPR criteria. Based on a medical advice, TPR or FPR could be emphasized by weighting its importance in Equation \ref{eq:RCR}.

Figure \ref{fig:CategoryResults} displays the RCR values obtained for the detection of cough with regard to each event category using the various sensors or their possible combination. The most ambiguous event classes turn out to be the throat clearing, forced expiration and laugh, for which a rather similar performance is carried out. For these three classes, the best results (achieved with all sensors) reach a RCR varying between 80 and 83\%. On the opposite, the distinction with the background and segments of speech leads to RCR values exceeding 95\%.

Regarding the potential of the various sensors, it is clearly observed that the chest belt yields the worst results, and this across all categories. On the other side, the audio microphone outperforms all other sensors in all conditions. For the four remaining sensors (ECG, thermistor, accelerometer and contact microphone), their relative ranking depends upon the considered event category. In this way, the accelerometer is noticed to be of high interest for the distinction with speech or with the background (for which abdominal movements are weak). The usefulness of the contact microphone is noticed for the discrimination with forced expirations and throat clearings. It is also worth noting that the thermistor is observed to be informative for the differentiation with laugh. This can be explained by the fact that laugh generally involves a nasal airflow \cite{BaSO01}, while the nose cavity is closed when coughing. Finally, the ECG is noticed to be ranked third or fourth sensor across all categories.

Inspecting results obtained with the combination of sensors, it is shown that the addition of other sensors to the audio microphone only leads to a minor improvement (and even sometimes to a slight reduction of performance), except for the distinction with speech and laugh. In these two latter cases, the complementarity mainly brought respectively by the accelerometer and the thermistor leads to an increase of RCR values of around 7 and 4\%.

The RCRs obtained by each sensor and their combination for the general cough detection problem (where all ambiguous event categories are mixed) are exhibited in Table \ref{tab:FrameResults}. As expected, the audio microphone and chest belt are clearly respectively the best and worst sensors with 93.2 and 67.1\%. Since the great majority of frames belong to the background and to speech (in a lesser extent), these results are mainly dominated by the performance for these two categories. It is therefore not surprising to find the accelerometer as second best sensor with about 84\%, followed by the contact microphone (79.6\%), the thermistor (76.3\%) and the ECG (72.7\%). Finally, it is emphasized again that the improvement brought by considering more sensors in complementarity with the audio microphone is negligible.

\begin{table} 
  \begin{center}
\vspace{0cm}
    \begin{tabular}{c|c|}
\textbf{Sensor} & \textbf{Revised Classification Rate (\%)}\\
\hline
\hline
ECG & 72.68\\
\hline
Thermistor & 76.34\\
\hline
Chest Belt & 67.1\\
\hline
Accelerometer & 83.94\\
\hline
Contact Microphone & 79.57\\
\hline
Audio Microphone & 93.17\\
\hline
Both Microphones & 93.22\\
\hline
Audio + Accelero & 93.27\\
\hline
Audio + Accelero + ECG & 92.49\\
\hline
All sensors & 92.9\\
\hline
    \end{tabular}
      \end{center}  
      \vspace{-.2cm}
  \caption{\emph{Revised Classification Rate at the frame level using various sensors and their possible combinations for the general cough detection problem.}}  
  \label{tab:FrameResults}
\vspace{-.8cm}
\end{table}

\section{Comparison with Karmelsonix}\label{sec:Karmelsonix}

The performance of the proposed approach is now compared to the Karmelsonix system, one of the few commercially available cough counters. Karmelsonix makes use of four sensors \cite{Vizel}: one audio microphone, two contact microphones (one on the trachea and one on the thorax), and a chest wall impedance belt. The acquisition system is also provided with an analysis software from which we here consider the cough detections.

In this experiment, training is performed on set A of the database (22 subjects) and the test is carried out on set B (10 subjects for which parallel recordings with Karmelsonix are available). The performance of the various sensors and their combination is assessed at the event level since Karmelsonix only detects cough event triggers. As standard metrics at the event level, we use the specificity and sensitivity measures.

In order to determine cough events with the proposed approach, our key idea was to focus only on the detection of the explosive phase of cough. Indeed it is known that the intermediate phase of cough is very similar to a forced expiration \cite{Korpas}. Besides, the voiced phase does not appear in all cough events and could be confused with some parts of laughing or throat clearing. On the opposite, the explosive phase is charateristic of the beginning of any cough event and is therefore distinctive enough from other ambiguous event categories. Since explosive phases have not been manually annotated, we consider in the remainder of this paper that they dominate the first 60 ms of the cough events.

\begin{figure}
  \begin{center}
   \includegraphics[width=8cm]{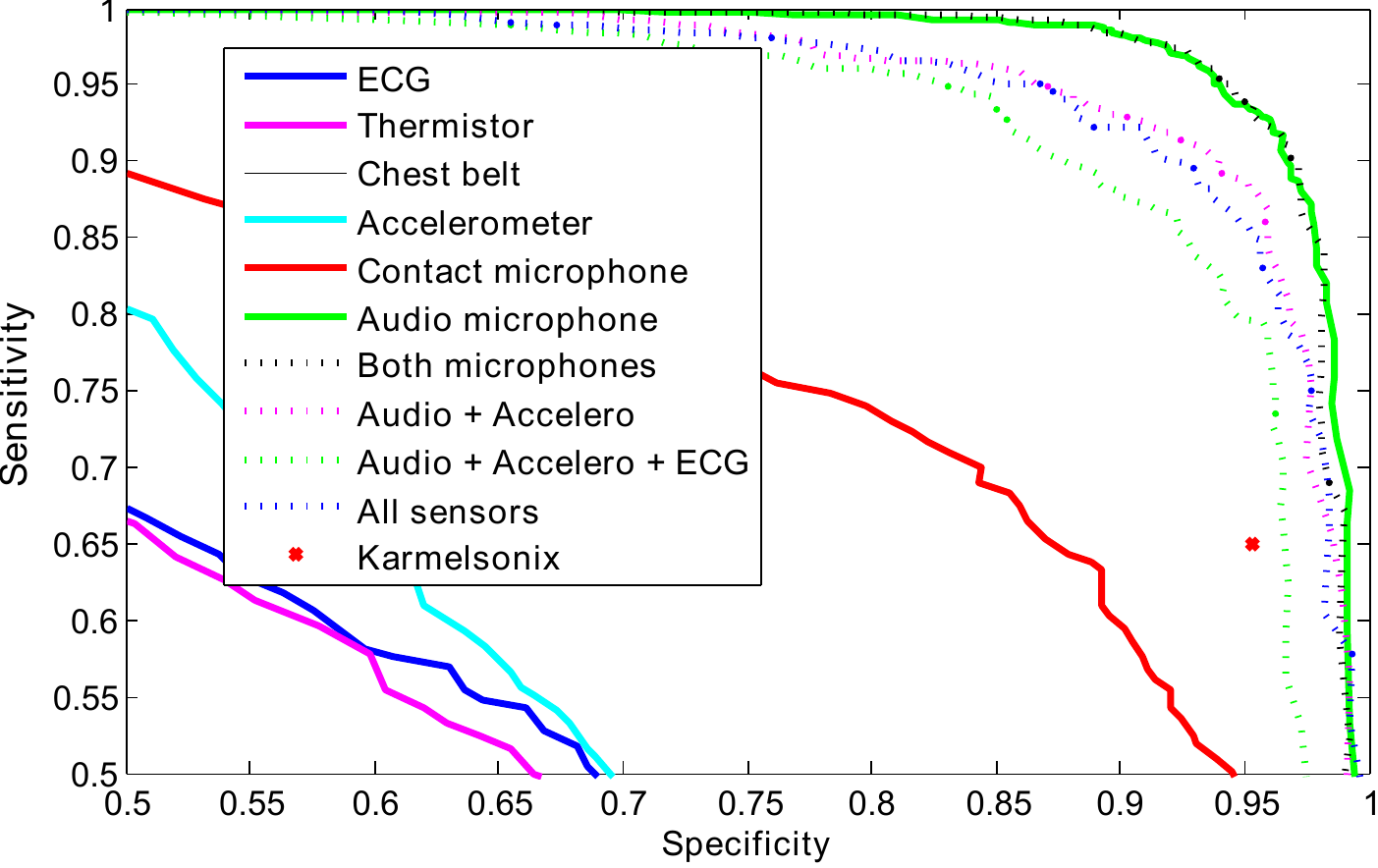}
  \end{center}
  \vspace{-.5cm}
\caption{{\it Performance of cough detection in the specificity-sensitivity plane, using several sensors and their possible combinations, and comparison with the Karmelsonix system.}} 
\vspace{-.2cm}
\label{fig:Event_FINAL}
\end{figure}

Figure \ref{fig:Event_FINAL} shows the results, in the specificity-sensitivity plane by varying the decision threshold, of the proposed approach based on each sensor separately or using their combination. The performance obtained by the Karmelsonix system is indicated as well. Note that the performance using the chest belt is so low that it is not represented in the plot range. It is also observed that considering the determination of the explosive phase for event detection modifies the relative ranking between the sensors, notably regarding the accelerometer and contact microphone. Nonetheless the audio microphone still clearly remains the most effective sensor. Furthermore our attempts to integrate a multimodal technique did not lead to any improvement compared to the audio-only approach. We therefore consider in the following that the proposed method is reduced to the use of the audio microphone. This approach is observed to outperform the Karmelsonix system which, albeit providing good specificity results (few false alarms), gives poor sensitivity results on average (about 65\%, indicating a high number of missed detections). Interestingly, the proposed approach is observed to provide, for a comparable specificity, much better sensitivity capabilities. As working point for the proposed technique, we fixed in the following the decision threshold such that a compromise between misses and false alarms is found (i.e. in order to have comparable specificity and sensitivity performance).

\begin{table} 
  \begin{center}
\vspace{0cm}
    \begin{tabular}{c||c|c||c|c|}
    
 & \multicolumn{2}{c||}{\bf{Proposed Audio-only}} & \multicolumn{2}{c|}{\bf{Karmelsonix}} \\
\hline 
\textbf{Subj.} & \textbf{Sens.(\%)} & \textbf{Spec.(\%)} & \textbf{Sens.(\%)} & \textbf{Spec.(\%)}\\
\hline 
1 & 88.9 & 100 & 93.1 & 100\\
\hline 
2 & 97.3 & 94.7 & 87.7 & 90.1\\
\hline 
3 & 79.2 & 96.6 & 0.0 & 100\\
\hline 
4 & 98.6 & 92.2 & 83.3 & 93.7\\
\hline 
5 & 100 & 88.9 & 61.4 & 86.0\\
\hline 
6 & 93.3 & 93.3 & 4.0 & 100\\
\hline 
7 & 98.6 & 98.6 & 68.1 & 92.4\\
\hline 
8 & 88.0 & 95.7 & 90.7 & 97.1\\
\hline 
9 & 100 & 91.1 & 70.4 & 96.2\\
\hline 
10 & 100 & 93.4 & 90.1 & 97.0\\
\hline 
\hline 
\textbf{Avg.} & \textbf{94.39} & \textbf{94.45} & \textbf{64.9} & \textbf{95.3}\\
\hline 
\textbf{STDV} & 7.0 & 3.4 & 34.8 & 4.7\\
\hline 
    \end{tabular}
  \end{center}
  \vspace{-.2cm}
  \caption{\emph{Detail of the performance achieved by the proposed approach using only the audio microphone and with the Karmelsonix system for the 10 subjects of set B.}}  
  \label{tab:CompKS}
\vspace{-.8cm}
\end{table}

The detection results specific to each subject are presented in Table \ref{tab:CompKS}. It is noted that Karmelsonix suffers from a wide inter-subject variability. Indeed, while subjects 1, 8 and 10 have a specificity higher than 90\%, it turns out that the system misses almost all cough events for subjects 3 and 6. This drawback is the most important inconvenient of Karmelsonix. Note that this observation is in contradiction with the preliminary study carried out by Karmelsonix developers in \cite{Vizel} and where they report an overall sensitivity of 96\%. On the opposite, while the proposed approach achieves comparable specificity results, its sensitivity never goes below 79\% and reaches an average value of about 94.5\%.

Finally, Table \ref{tab:Confusion} highlights the sources of false alarms for the proposed audio-based technique and the Karmelsonix system. Segments of speech and background noise are noticed to be well discriminated, with 6 false alarms for the proposed method and 2 for Karmelsonix, over a duration of 58 minutes. For the proposed approach, the most confusing classes are respectively laugh (18.7\%), throat clearing (8.8\%) and forced expiration (5.5\%). Regarding the ambiguity for the Karmelsonix system, it turns out that throat clearing leads to the highest confusion (16.2\%) while forced expiration and laugh are well discriminated. Albeit the proposed audio-based method leads to a higher number of false alarms (42 against 28 for Karmelsonix), it also produces much less missed cough events, providing on average a considerable improvement. Indeed the Karmelsonix system is clearly seen to miss an important amount of cough events across all cough types: 28.1\% for fits of coughing, around 36\% for cough with high and intermediate volume, and 44.5\% for cough with low volume. On the opposite, the proposed approach shows up a low miss rate across all cough types (ranging from 8.7\% on cough with an intermediate volume, to 0\% for cough with high volume).

\begin{table} 
  \begin{center}
\vspace{0cm}
    \begin{tabular}{c|c|c|c|}
 &  & \bf{\# Cough detected} & \bf{\# Cough detected}\\
 & & \bf{with the} & \bf{with}\\
 & \bf{Amount} & \bf{proposed system} & \bf{Karmelsonix}\\
\hline 
\hline 
\bf{High vol. cough} & 148 & 148 & 95\\
\hline 
\bf{Int. vol. cough} & 150 & 137 & 95\\
\hline 
\bf{Low vol. cough} & 155 & 148 & 86\\
\hline 
\bf{Fit of coughing} & 267 & 251 & 192\\
\hline 
\bf{Expiration} & 90 & 5 & 1 \\
\hline 
\bf{Throat Clearing} & 148 & 13 & 24 \\
\hline 
\bf{Speech} & 474 sec & 1 & 0\\
\hline 
\bf{Laugh} & 96 & 18 & 1 \\
\hline 
\bf{Background} & 3006 sec & 5 & 2\\ 
\hline 
    \end{tabular}
  \end{center}
  \vspace{-.2cm}
  \caption{\emph{Repartition of cough events detected by the proposed audio-based approach and the Karmelsonix system across all event categories.}}
  \label{tab:Confusion}
\vspace{-.8cm}
\end{table}

\section{Conclusion and Discussion}\label{sec:Conclusion}
This paper aimed at providing an objective study on the relevance of several sensors to use in a cough detection system: ECG, thermistor, chest belt, accelerometer, contact and audio mircophones. In a first experiment features extracted for each sensor were assessed through the mutual information they convey for the classification problem. Already at this stage it was highlighted that the audio microphone is the most promising sensor while the chest belt is the worst one. In a second time we compared the ability of sensors for distinguish cough from several other event categories: background, forced expiration, throat clearing, speech and laugh. Following an analysis at the frame level, it turned out that considering other sensors in addition to the audio microphone did not provide any substantial improvement, except regarding the discrimination of cough with speech (for which the accelerometer was observed to be particularly useful) and with laugh (where the thermistor showed an interesting performance). Regarding the most difficult categories to differentiate, it was seen that forced expiration, throat clearing (particularly abundantly produced by subjects with cystic fibrosis) and laugh both led to the highest ambiguity with a misclassification rate of about 20\% for all. Thirdly and lastly, cough event detection was carried out by recognizing the explosive phase of cough, and the proposed approach was compared to the Karmelsonix system, one of the few commercially available solutions. Two important conclusions could be drawn from this experiment. First, the audio microphone achieved the best results, even when compared to our attempts of multimodal approaches. This can be explained by the fact that the acoustic properties during the explosive phase of cough are sufficiently well differentiated and that considering other modalities does not bring any improvement for detecting cough event via this framework. From a practical point of view, this is a strong advantage since the cough monitor can be reduced to a simple audio microphone which is discreete and suited for an ambulatory 24-hour usage. The second conclusion was that the performance yielded by the proposed approach reached around 94.5\% of both specificity and sensitivity, clearly outpeforming the Karmelsonix system. Karmelsonix was observed to suffer from an important inter-subject variability, with a sensitivity which coud be very low for some subjects. On the opposite, this drawback was alleviated in our proposed audio-only method whose sensitivity varied between 79\% and 100\%, for a specificity ranging from 89\% to 100\%. Our future works encompass the validation of our prototype in patients with cystic fibrosis (and possibly other chronic cough diseases), in a 24-hour ambulatory context, as well as the qualification of cough events to complement their frequency quantification.

\section*{Acknowledgment}
The project is supported by the Walloon Region (Grant WIST 3 COMPTOUX \# 1017071).

\ifCLASSOPTIONcaptionsoff
  \newpage
\fi

%
%
%

\end{document}